\documentclass{PoS}

\title{Long term variability of the blazar PKS 2155-304}

\ShortTitle{Long term variability of the blazar PKS 2155-304}

\author{\speaker{Jill Chevalier}\\
        Laboratoire d'Annecy-le-Vieux de Physique des Particules, Université Savoie Mont-Blanc, CNRS/IN2P3, F-74941 Annecy-le-Vieux, France\\
        E-mail: \email{jill.chevalier@lapp.in2p3.fr}}

\author{Max Anton Kastendieck\\
        Universität Hamburg, Institut für Experimentalphysik, Luruper Chaussee 149, D 22761 Hamburg, Germany}
\author{Frank Rieger\\
        Max-Planck-Institut für Kernphysik, P.O. Box 103980, D 69029 Heidelberg, Germany}
\author{Gilles Maurin\\
        Laboratoire d'Annecy-le-Vieux de Physique des Particules, Université Savoie Mont-Blanc, CNRS/IN2P3, F-74941 Annecy-le-Vieux, France}
\author{Jean-Philippe Lenain\\
        LPNHE, Université Pierre et Marie Curie Paris 6, Université Denis Diderot Paris 7, CNRS/IN2P3, 4 Place Jussieu, F-75252, Paris Cedex 5, France}
\author{Giovanni Lamanna\\
        Laboratoire d'Annecy-le-Vieux de Physique des Particules, Université Savoie Mont-Blanc, CNRS/IN2P3, F-74941 Annecy-le-Vieux, France}
\author{for the H.E.S.S. collaboration}
                                
\abstract{Time variability of the photon flux is a known feature of active galactic nuclei (AGN) and in particular of blazars. The high frequency peaked BL Lac (HBL) object PKS~2155-304 is one of the brightest sources in the TeV band and has been monitored regularly with different instruments and in particular with the H.E.S.S. experiment above 200 GeV for more than 11 years. These data together with the observations of other instruments and monitoring programs like SMARTS (optical), \textit{Swift}-XRT/\textit{RXTE}/XMM-Newton (X-ray) and \textit{Fermi}-LAT (100 MeV < E < 300 GeV) are used to characterize the variability of this object in the quiescent state over a wide energy range. Variability studies are made by looking at the lognormality of the light curves and at the fractional root mean square (rms) variability $F_{\rm var}$ in several energy bands. 
Lognormality is found in every energy range and the evolution of  $F_{\rm var}$ with the energy shows a similar increase both in X-rays and in TeV bands.}

\FullConference{The 34th International Cosmic Ray Conference,\\
		30 July- 6 August, 2015\\
		The Hague, The Netherlands}

\begin{document}

\section{Introduction}

As many blazars, PKS 2155-304 (z = 0.117) features a questioning and interesting behavior: its variability which can be seen over a great energy range, from radio to TeV. Characterizing the variability of blazars is important to study the underlying acceleration processes which are, at the moment, poorly known. Multiwavelength (MWL) data, from experiments and instruments as H.E.S.S., \textit{Fermi}-LAT, \textit{RXTE}, \textit{Swift}-XRT, XMM-Newton and SMARTS are used to study the variability of this object. 

Lognormality and intrinsic variability of the MWL light curves are investigated.
Lognormal flux variability was first found in galactic accreting sources like X-ray binaries, linking lognormal processes to accretion disks \cite{bib:Uttley2001}. In a lognormal process, the fluctuations of the flux are on average proportional, or at least correlated, to the flux itself, ruling out additive processes in favor of multiplicative ones. For blazars, a lognormal flux behaviour could as well be indicative of the accretion disk variability's imprint onto the jet \cite{bib:McHardy2008}. 
BL Lacertae was the first blazar to show a lognormal behavior in the X-ray \cite{bib:Giebels2009}.
For PKS~2155-304, the first detection of a lognormal behavior was during the flaring state of 2006 in H.E.S.S. data \cite{bib:HESS2010}.

Studying the evolution of the variability with the energy is one element to further distinguish between different acceleration models. 


\begin{figure}[t]
\begin{center}
\includegraphics[width=1.0\textwidth]{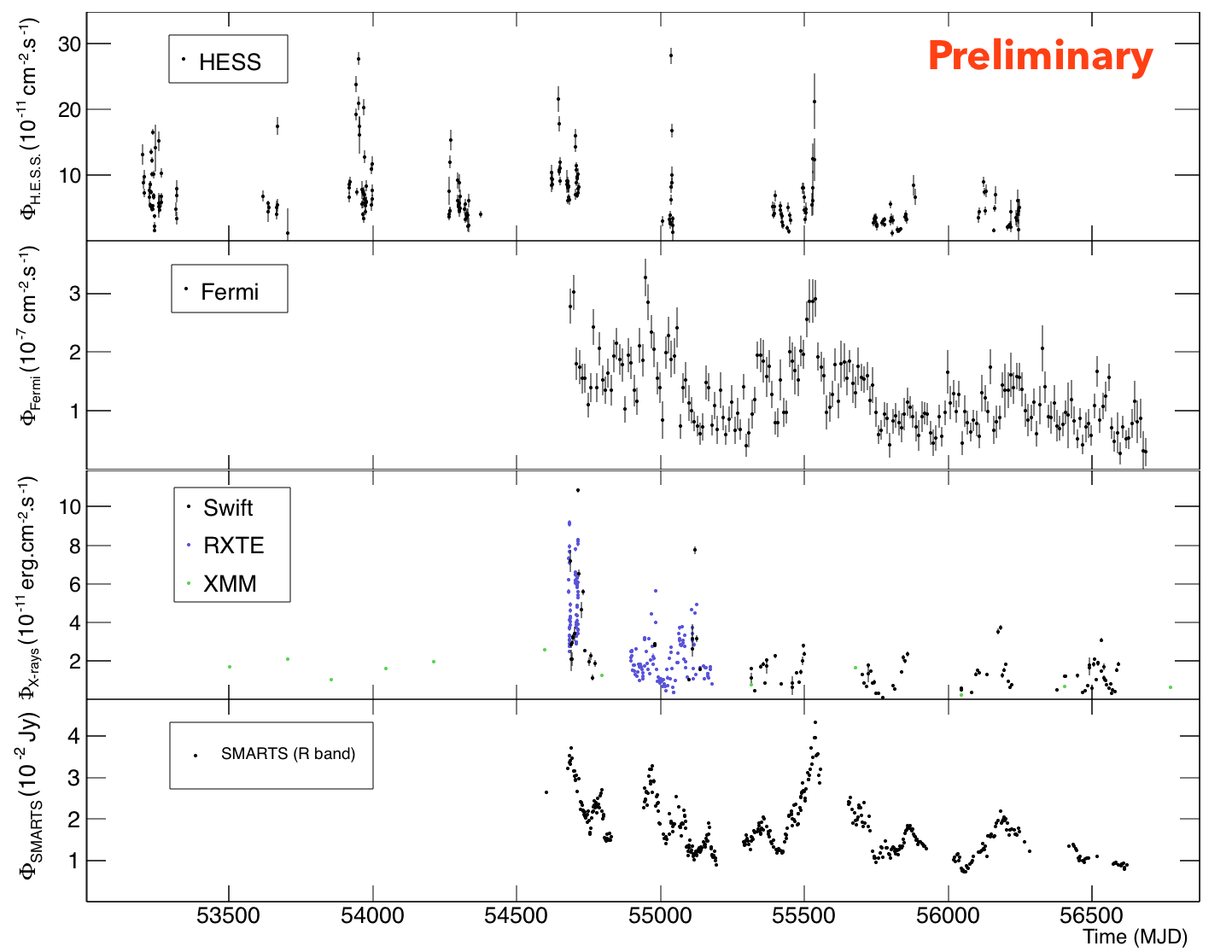}
\caption{Light curves from top to bottom : H.E.S.S. in ${\rm cm}^{-2}.{\rm s}^{-1}$ from 0.2 to 125 TeV, \textit{Fermi}-LAT in ${\rm cm}^{-2}.{\rm s}^{-1}$ from 0.1 to 300 GeV, X-ray in ${\rm erg}.{\rm cm}^{-2}.{\rm s}^{-1}$ from 2 to 10 keV (\textit{Swift}-XRT in black, \textit{RXTE} in blue and XMM-Newton in green) and SMARTS (R band) in Jy.}
\label{plot:LC}
\end{center}
\end{figure}

\section{Data sets}
\label{section:data}

\subsection{Very Hight Energy (VHE): H.E.S.S.}

The H.E.S.S. (High Energy Stereoscopic System) experiment is an array of 5 ground-based Cherenkov Telescopes which detect the Cherenkov light emitted by the shower created by VHE $\gamma$-rays with energies from $\sim100$ GeV to $>10$ TeV. The four first telescopes are operated since 2003 (HESS-I phase) and a fifth telescope had been added to the array in 2012 (HESS-II phase).

The present analysis is based on the data taken between MJD 53200 and 56246 using only the HESS-I phase of the array (without the fifth telescope) collected with 3 or 4 telescopes. The flaring state of July/August 2006 from MJD 53943 to 53949 has been removed, in order to have a light curve of the quiescent state of PKS~2155-304 \cite{bib:HESS2010}. In total, about 260h of data passed standard quality cuts as defined in \cite{bib:HESS2006}, with a mean zenith angle of $21^{\circ}$ and an energy threshold of 180GeV. The data are analyzed with the Model analysis chain (Model++) \cite{bib:Naurois2009} between 0.2 and 125 TeV, yielding to a total detection significance $> 300 \sigma$. 

The light curve of nightly fluxes used here is calculated assuming a power-law spectrum $dN/dE \propto E^{-\Gamma}$, $\Gamma$ being the power law index. However, it has been reported in \cite{bib:HESS2010} that PKS~2155-304 shows some spectral variability. Therefore the fluxes are derived with a different fitted spectral index each year to take into account this variability. This leads to 9 different values for $\Gamma$ (one per year) with a mean of $3.45 \pm 0.05$  and a standard deviation of 3\%.

The resulting quiescent nightly light curve, presented in the first panel of figure \ref{plot:LC}, has a mean flux of $6.9 \times 10^{-11} {\rm ph.cm}^{-2}.{\rm s}^{-1}$. This analysis procedure has been applied to the same data set using the following energy cuts: 0.2-0.63 and 0.63-125 TeV, the cut at 0.63 TeV being chosen to have the same statistics in each energy band.

\subsection{High Energy (HE): \textit{Fermi}-LAT}

The \textit{Fermi} Large Area Telescope (LAT) \cite{bib:Atwood2009} on board the \textit{Fermi} satellite is a pair-conversion telescope designed to detect HE $\gamma$-rays in the energy range from $\sim 20$ MeV to more than 500 GeV. 

The data for PKS~2155-304 used here 
goes from MJD 54687 to 56687 and are taken between 100 MeV and 300 GeV. The detector is described by the P7REP\_SOURCE\_V15 instrument response functions\footnote{http://fermi.gsfc.nasa.gov/ssc/data/analysis/documentation/Cicerone/Cicerone\_Data\_Exploration/} and the analysis has been performed by the tool enrico \cite{bib:Sanchez2013} using \textit{Fermi} Science Tools v9r32p5. The prefactor of the diffuse galactic background and the normalisation of the isotropic diffuse emission are left free to vary in the likelihood fit. All sources from the 3FGL catalog \cite{bib:Nolan2012} within $15^{\circ}$ around PKS~2155-304 are included in the model. The spectra of PKS~2155-304 and the nearby FSRQ (Flat Spectrum Radio Quasar) PKS~2149-306 are modeled by a simple power-law, and their spectral indices and prefactors are left free. All other components are fixed to the values in the 3FGL catalog. 

PKS~2155-304 is detected with a significance of $> 150 \sigma$ with an overall spectral index of $1.83 \pm 0.01$. The photon counts are integrated into 201 bins of 10 days, resulting in a light curve, presented in the second panel of figure \ref{plot:LC}, with a mean flux of $1.3 \times 10^{-7} {\rm ph.cm}^{-2}.{\rm s}^{-1}$. The same analysis procedure has been applied to the same data set using the following energy cuts : 0.1-3, 3-30 and 30-300 GeV.

\subsection{X-ray: \textit{RXTE}, \textit{Swift}-XRT and XMM-Newton}

\textit{RXTE} (\textit{Rossi X-ray Timing Explorer}) already analyzed light curves are publicly available\footnote{http://cass.ucsd.edu/~rxteagn/} between 2 and 10 keV, and in three other energy bands: 2-4, 4-7 and 7-10 keV. 

\textit{Swift} X-ray Telescope (XRT) data were analyzed using the package \texttt{HEASOFT 6.16} between 2 and 10 keV, and also in four other energy bands: 0.3-2, 2-4, 4-7 and 7-10 keV. The data were recalibrated using the last update of \texttt{CALDB} and reduced using the standard procedures \texttt{xrtpipeline} and \texttt{xrtproducts}. Caution has been taken to properly account for pile-up effects for corresponding affected exposures, and spectral fits were performed using \texttt{Xspec 12.8.2} assuming a power-law spectrum.

XMM-Newton (X-ray Multi Mirror Mission) public EPIC (European Photon Imaging Camera) data were reduced using the SAS software package, version 14.0, and analyzed following \cite{bib:Tatischeff2012} between 2 and 10 keV, and also in four other energy bands: 0.3-2, 2-4, 4-7 and 7-10 keV, to match \textit{RXTE} and \textit{Swift}-XRT data. 

\textit{Swift}-XRT and XMM-Newton data have been corrected from galactic absorption with $n_{\rm H} = 1.48 \times 10^{20} {\rm cm}^{-2}$ \cite{LABsurvey}.

In all common energy bins, data of the three instruments are combined to construct an all-over X-ray light curve. In order to have an X-ray light curve of the quiescent state of PKS~2155-304, all ToO (Target of Opportunities) data were removed. The final light curve, presented in the third panel of figure \ref{plot:LC}, goes from MJD 53501 to 55474 with a mean flux of $2.7 \times 10^{-11} {\rm erg.cm}^{-2}{\rm .s}^{-1}$.

\subsection{Optical wavelengths : SMARTS}

SMARTS (Small and Moderate Aperture Research Telescope System) already analyzed data are publicly available\footnote{http://www.astro.yale.edu/smarts/glast/home.php\#}. Magnitudes are corrected for the absorption of the galactic foreground using \cite{bib:SF11} before being converted to a spectral flux density using zero magnitude flux values of \cite{bib:Cohen1992}. The light curves are taken in the J, R, V and B bands from MJD 54603 to 56622. For the sake of simplicity, only the light curve for the R band is presented in the last panel of figure \ref{plot:LC}.

\section{Results}

\subsection{Lognormality}
\label{section:lognormality}

\begin{table}[h]
\begin{center}
\begin{tabular}{l c c c c }
\hline
\hline
  & \multicolumn{2}{c}{$\Phi$} & \multicolumn{2}{c}{log $\Phi$} \\
 & $\chi^2/$d.o.f. & Prob & $\chi^2/$d.o.f. & Prob \\
 \hline
  H.E.S.S. (> 200GeV) & 2.71 & $10^{-4}$ & 0.716 & 0.75 \\
  \textit{Fermi}-LAT (0.1-300GeV) & 2.51 & $10^{-3}$ & 1.36 & 0.18 \\
  X-ray (2-10keV) & 6.67 & $10^{-10}$ & 1.36 & 0.20 \\
  SMARTS (B band) & 4.35 & $10^{-3}$ & 0.422 & 0.96 \\
  SMARTS (V band) & 2.30 & $10^{-9}$ & 0.790 & 0.65 \\
  SMARTS (R band) & 5.42 & $10^{-9}$ & 0.823 & 0.62 \\
   SMARTS (J band) & 2.33 & $10^{-3}$ & 1.28 & 0.22 \\
\hline
\hline
\end{tabular}
\caption{Values of the $\chi^2$ and associated probability (probability that the model is good) for the Gaussian fit of the flux  and log flux distributions for each light curve.}
\label{table:histo}
\end{center}
\end{table}

The lognormal behavior of PKS~2155-304 is investigated by looking at the distribution of fluxes and the correlation between the excess rms and the average flux. 

The flux and log flux distributions of each light curve are fitted by a Gaussian using a $\chi^2$ fit. The results of the fits are summarized in table \ref{table:histo}. In all cases, a Gaussian fits the log flux distribution better than the flux distribution, indicating that MLW fluxes of PKS~2155-304 follow a lognormal distribution.

\begin{figure*}[h!]
\begin{center}
\includegraphics[width=0.45\textwidth]{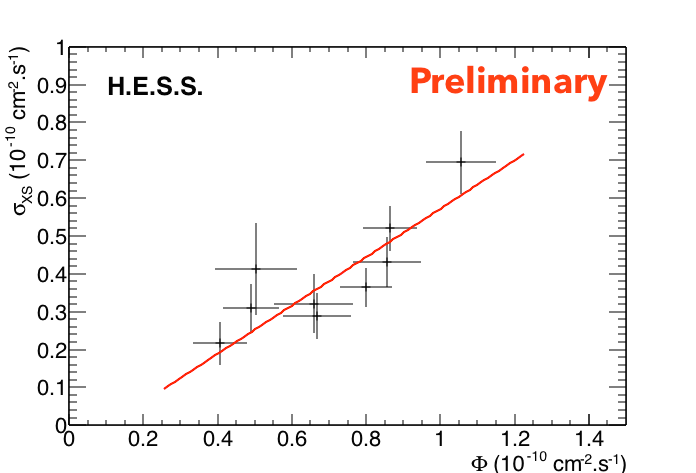}
\includegraphics[width=0.45\textwidth]{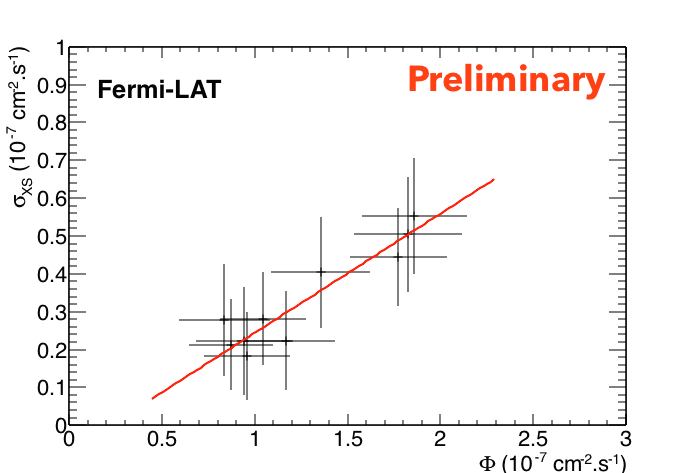}
\includegraphics[width=0.45\textwidth]{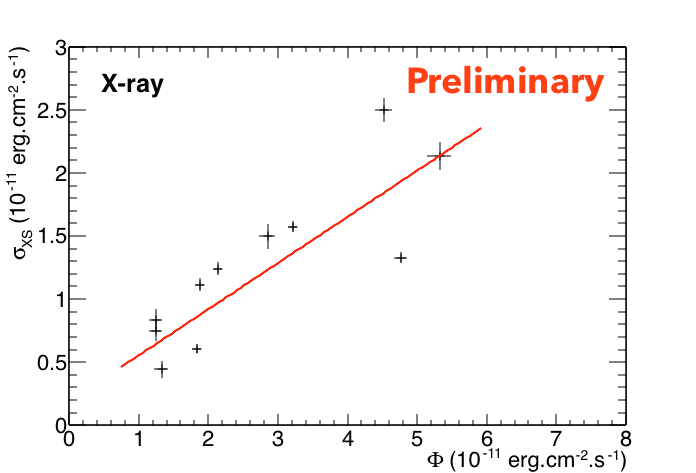}
\includegraphics[width=0.45\textwidth]{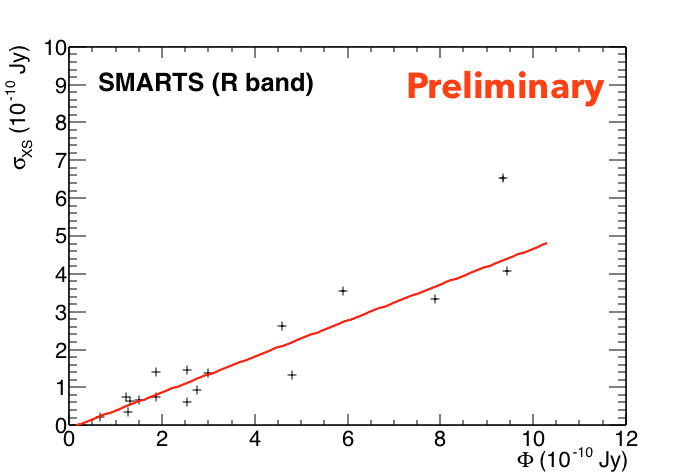}
\caption{Scatter plot of the excess rms and the average flux for the MWL data.
A linear fit is shown in red.}
\label{plot:excess}
\end{center}
\end{figure*}

The scatter plots of the excess rms $\sigma_{\rm XS} = \sqrt{S^2 - \overline{\sigma_{\rm err}}}$, $S^2$ being the variance and $ \overline{\sigma_{\rm err}^2}$ the mean square error of the light curve \cite{bib:Vaughan2003}, versus the mean flux $\overline{\Phi}$ are shown for each light curve in figure \ref{plot:excess}. Each point is calculated with at least 20 light curve points.
To test the rms-flux relation, scatter plots are fitted by a constant and a linear adjustments. The fit results are summarized in table \ref{table:excess}. The results show more than a $5 \sigma$ preference for the linear fit except for the \textit{Fermi}-LAT data. However the correlation between $\sigma_{\rm XS}$ and $\overline{\Phi}$ is not expected to be linear, especially when the points are spread in the plot, 
so the correlation factor $\rho$ and the Kendall rank $\tau$ (which measure the ordering of the points \cite{bib:Gleissner04}) are calculated. In all cases $\rho > 0.80$ and $\tau > 0.65$ meaning that $\sigma_{\rm XS}$ and $\overline{\Phi}$ show a strong correlation, leading to the conclusion that the fluctuations of the flux are correlated and/or proportional to the flux.

\begin{table}[h!]
\begin{center}
\begin{tabular}{l c c c c c}
\hline
\hline
  & cst & cst$+ a \times x$ &  &  \\
 & $\chi^2/$d.o.f. & $\chi^2/$d.o.f. & $\sigma$ & $\rho$ & $\tau$ \\
 \hline
  H.E.S.S. (> 200GeV) & 4.15 & 0.773 & 5.25 &$0.84 \pm 0.15$ & $0.72 \pm 0.27$  \\
  \textit{Fermi}-LAT (0.1-300GeV) & 0.937 & 0.107 & 2.75 & $0.93 \pm 0.19$ & $0.69 \pm 0.25$\\
  X-ray (2-10keV) & 78.2 & 28.9 & 23 & $0.85 \pm 0.02$ & $0.67 \pm 0.24$  \\
  SMARTS (B band) & 4222 & 892 & 225 & $0.92 \pm 0.03$ & $0.68 \pm 0.18$ \\
  SMARTS (V band) & 4472 & 810 & 236 & $0.92 \pm 0.02$ & $0.75 \pm 0.19$ \\
  SMARTS (R band) & 7542 & 946 & 326 & $0.93 \pm 0.02$ & $0.74 \pm 0.18$ \\
  SMARTS (J band) & 3339 & 398  & 197 & $0.94 \pm 0.03$ & $0.80 \pm 0.21$ \\
\hline
\hline
\end{tabular}
\caption{Values of the reduced $\chi^2$ of the constant and linear fits of the scatter plots shown in figure \protect \ref{plot:excess} for each light curve, with values for the significance $\sigma$, the correlation factor $\rho$ and the Kendall rank 
$\tau$.}
\label{table:excess}
\end{center}
\end{table}

\subsection{Variability energy distribution (VED)}
\label{section:fvar}

The fractional variability $F_{\rm var}$ is used to quantify the variability of the flux taking into account the errors \cite{bib:Vaughan2003}:
$$ F_{\rm var} = \sqrt{{S^2 - \overline{\sigma^2_{\rm err}} \over \overline{\Phi}^2}} \pm \sqrt{\left( \sqrt{{1 \over 2N}} { \overline{\sigma^2_{\rm err}} \over \overline{\Phi}^2 F_{\rm var}}  \right)^2 + \left( \sqrt{\overline{\sigma^2_{\rm err}} \over N} {1 \over \overline{\Phi}} \right)^2} $$

\begin{figure}[h!]
\begin{center}
\includegraphics[width=1\textwidth]{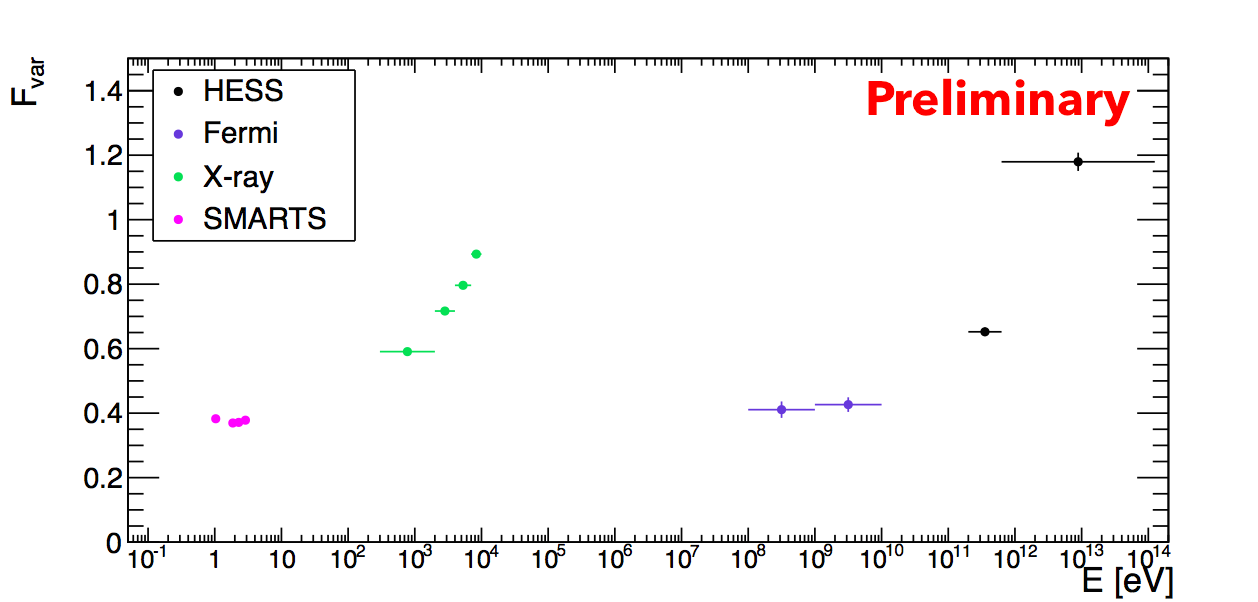}
\caption{VED of PKS~2155-304. } 
\label{plot:Fvar}
\end{center}
\end{figure}

It is computed for each data set at the different energies presented in section \ref{section:data}. The VED plot is shown on figure \ref{plot:Fvar}. This figures shows that the SMARTS variability level is somewhat constant while a clear increase is visible in the X-ray band. Then the variability is smaller in the \textit{Fermi}-LAT range and increases again in the H.E.S.S. range. 


\section{Conclusion}

The distribution of the flux of the quiescent state of PKS~2155-304 follows a lognormal distribution and the variations of the flux are correlated with the flux level itself in all energy band probed, indicating that lognormality is an instrinsic characteristic of PKS~2155-304. This results have already been shown for the flaring state of 2006 in the TeV band \cite{bib:HESS2010}. Others BL Lac type sources have present the same lognormal behavior, like BL Lacertae in the X-ray range \cite{bib:Giebels2009} and Markarian 501 at TeV energies \cite{bib:Chakraborty2015}. Interestingly non BL Lac objects such as the Seyfert~1 galaxy IRAS~13~244-3809 have revealed a similar behavior \cite{bib:Gaskell2004}. Seyfert~1 galaxies have different characteristics compare to BL Lac: radio quiet (no jet domination), with emission lines and more gas surrounding the central engine, indicating again that the lognormal behavior is disk related.

Moreover, the VED of PKS~2155-304 is showing an interesting feature with a similar variability increase in the X-ray and TeV ranges. 
Such a shape of the VED could help distinguish between different acceleration models in blazars as they won't give the same variability signature.

\bigskip 

\footnotesize{{\bf Acknowledgment: }{The support of the Namibian authorities and of the University
of Namibia in facilitating the construction and operation of H.E.S.S. is gratefully
acknowledged, as is the support by the German Ministry for Education and
Research (BMBF), the Max Planck Society, the German Research Foundation
(DFG), the French Ministry for Research, the CNRS-IN2P3 and the Astroparticle
Interdisciplinary Programme of the CNRS, the U.K. Science and Technology
Facilities Council (STFC), the IPNP of the Charles University, the Czech Science
Foundation, the Polish Ministry of Science and Higher Education, the South
African Department of Science and Technology and National Research Foundation,
and by the University of Namibia. We appreciate the excellent work of the
technical support staff in Berlin, Durham, Hamburg, Heidelberg, Palaiseau, Paris,
Saclay, and in Namibia in the construction and operation of the equipment. 
This paper has made use of up-to-date SMARTS optical/near-infrared light curves that are available at www.astro.yale.edu/smarts/glast/home.php.}}


\begin{thebibliography}{99}

\bibitem{bib:Uttley2001} Uttley, P., \& McHardy, I. M. (2001). \textit{The flux-dependent amplitude of broadband noise variability in X-ray binaries and active galaxies}. Monthly Notices of the Royal Astronomical Society, 323(2), L26-L30 [{\tt arXiv:astro-ph/0103367}].

\bibitem{bib:McHardy2008} McHardy, I., \textit{Explaining X-ray variability in blazars}, \textit{Workshop on Blazar Variability across the Electromagnetic Spectrum} (2008). 

\bibitem{bib:Giebels2009} Giebels, B., \& Degrange, B. (2009). \textit{Lognormal variability in BL Lacertae}. Astronomy \& Astrophysics, 503(3), 797-799 [{\tt arXiv:0907.2425}].

\bibitem{bib:HESS2010} Abramowski, A., Acero, F., Aharonian, F., Akhperjanian, A. G., Anton, G., De Almeida, U. B., ... \& Fegan, S. (2010). \textit{VHE $\gamma$-ray emission of PKS 2155-304: spectral and temporal variability}. Astronomy \& Astrophysics, 520, A83 [{\tt arXiv:1005.3702}].

\bibitem{bib:HESS2006} Aharonian, F., Akhperjanian, A. G., Bazer-Bachi, A. R., Beilicke, M., Benbow, W., Berge, D., ... \& De Jager, O. C. (2006). \textit{Observations of the Crab nebula with HESS}. Astronomy \& Astrophysics, 457(3), 899-915 [{\tt arXiv:astro-ph/0607333}].

\bibitem{bib:Naurois2009} De Naurois, M., \& Rolland, L. (2009). \textit{A high performance likelihood reconstruction of $\gamma$-rays for imaging atmospheric Cherenkov telescopes}. Astroparticle Physics, 32(5), 231-252 [{\tt arXiv:0907.2610}].

\bibitem{bib:Atwood2009} Atwood, W. B., Abdo, A. A., Ackermann, M., Althouse, W., Anderson, B., Axelsson, M., ... \& Ciprini, S. (2009). \textit{The Large Area Telescope on the Fermi gamma-ray space telescope mission}. The Astrophysical Journal, 697(2), 1071 [{\tt arXiv:0902.1089}].

\bibitem{bib:Sanchez2013} Sanchez, D. A., \& Deil, C. (2013), \textit{Enrico : a Python package to simplify Fermi-LAT analysis} [{\tt arXiv:1307.4534}].

\bibitem{bib:Nolan2012} Acero, F., Ackermann, M., Ajello, M., Albert, A., Atwood, W. B., Axelsson, M., ... \& Di Venere, L. (2015). \textit{Fermi Large Area Telescope Third Source Catalog}. The Astrophysical Journal Supplement Series, 218(2), 23 [{\tt arXiv:1501.02003}].

\bibitem{bib:Tatischeff2012} Tatischeff, V., Decourchelle, A., \& Maurin, G. (2012). \textit{Nonthermal X-rays from low-energy cosmic rays: application to the 6.4 keV line emission from the Arches cluster region}. Astronomy \& Astrophysics, 546, A88 [{\tt arXiv:1210.2108}].

\bibitem{LABsurvey} Kalberla, P.M.W., Burton, W.B., Hartmann, Dap, Arnal, E.M., Bajaja, E., Morras, R., \& Poppel, W.G.L. (2005).  \textit{The Leiden/Argentine/Bonn (LAB) Survey of Galactic HI, Final data release of the combined LDS and IAR surveys with improved stray-radiation corrections}. Astronomy \& Astrophysics, 440, 775 [{\tt arXiv:astro-ph/0504140}]

\bibitem{bib:SF11} Schlafly, E. F., \& Finkbeiner, D. P. (2011). \textit{Measuring reddening with Sloan Digital Sky Survey stellar spectra and recalibrating SFD}. The Astrophysical Journal, 737(2), 103 [{\tt arXiv:1012.4804}].

\bibitem{bib:Cohen1992} Cohen, M., Walker, R. G., Barlow, M. J., \& Deacon, J. R. (1992). \textit{Spectral irradiance calibration in the infrared. I-Ground-based and IRAS broadband calibrations}. The Astronomical Journal, 104, 1650-1657.

\bibitem{bib:Gleissner04} Gleissner, T., Wilms, J., Pottschmidt, K., Uttley, P., Nowak, M. A., \& Staubert, R. (2004). \textit{Long term variability of Cyg X-1-II. The rms-flux relation}. Astronomy \& Astrophysics, 414(3), 1091-1104 [{\tt arXiv:astro-ph/0311039}].

\bibitem{bib:Vaughan2003} Vaughan, S., Edelson, R., Warwick, R. S., \& Uttley, P. (2003). \textit{On characterizing the variability properties of X-ray light curves from active galaxies}. Monthly Notices of the Royal Astronomical Society, 345(4), 1271-1284 [{\tt arXiv:astro-ph/0307420}].

\bibitem{bib:Chakraborty2015} Chakraborty, N., Cologna, G., Rieger, F., et al. for the H.E.S.S. Collaboration, \textit{Rapid variability at very high energy in Mrk 501}, \textit{Proceedings of the 34th International Cosmic Ray Conference (ICRC 2015), The Hague (Netherlands)} (2015).

\bibitem{bib:Gaskell2004} Gaskell, C. M. (2004). \textit{Lognormal X-ray flux variations in an extreme narrow-line Seyfert~1 galaxy}. The Astrophysical Journal Letters, 612(1), L21.

\end{thebibliography}
\end{document}